\documentclass[conference]{IEEEtran}
\IEEEoverridecommandlockouts

\usepackage{cite}
\usepackage{amsmath,amssymb,amsfonts}
\usepackage{algorithmic}
\usepackage{graphicx}
\usepackage{textcomp}
\usepackage{xcolor}
\usepackage{comment}
\usepackage{listings}
\usepackage{fancyvrb}
\usepackage{framed}
\usepackage[utf8]{inputenc}
\usepackage{hyperref}
\UseRawInputEncoding

\usepackage{tabularx,booktabs}
\newcolumntype{C}{>{\centering\arraybackslash}X} 
\usepackage[listings,skins]{tcolorbox}
\usepackage[skipbelow=\topskip,skipabove=\topskip]{mdframed}
\def\BibTeX{{\rm B\kern-.05em{\sc i\kern-.025em b}\kern-.08em
    T\kern-.1667em\lower.7ex\hbox{E}\kern-.125emX}}
\begin{document}
\setlength{\columnsep}{0.2 in}

\title{Ensemble Defense System: A Hybrid IDS Approach for Effective Cyber Threat Detection}

\author{\IEEEauthorblockN{Sarah Alharbi}
\IEEEauthorblockA{\textit{Department of Electrical and Computer Engineering} \\
\textit{University of Delaware}\\
Newark, DE \\
sarahalh@udel.edu}
\and
\IEEEauthorblockN{Arshiya Khan}
\IEEEauthorblockA{\textit{Department of Electrical and Computer Engineering} \\
\textit{University of Delaware}\\
Newark, DE \\
arshiyak@udel.edu}
}
\maketitle

\begin{abstract}
Sophisticated cyber attacks present significant challenges for organizations in detecting and preventing such threats. To address this critical need for advanced defense mechanisms, we propose an Ensemble Defense System (EDS). An EDS is a cybersecurity framework aggregating multiple security tools designed to monitor and alert an organization during cyber attacks. The proposed EDS leverages a comprehensive range of Intrusion Detection System (IDS) capabilities by introducing a hybrid of signature-based IDS and anomaly-based IDS tools. It also incorporates Elasticsearch, an open-source Security Information and Event Management (SIEM) tool, to facilitate data analysis and interactive visualization of alerts generated from IDSs. The effectiveness of the EDS is evaluated through a payload from a bash script that executes various attacks, including port scanning, privilege escalation, and Denial-of-Service (DoS). The evaluation demonstrates the EDS's ability to detect diverse cyber attacks.
\end{abstract}


\section{Introduction}
Organizations today are encountered with threats from cybercriminals who continuously develop sophisticated attack techniques, targeting vulnerabilities within network systems and compromising sensitive data. Consequently, robust and thorough defense mechanisms have become imperative to ensure organizations' security in the face of these threats. The Ensemble Defense System (EDS) is at the forefront of cybersecurity strategies. EDS is a security implementation strategy that uses multiple layers of defense mechanisms and security tools to protect a network of machines from multifarious threats and attacks \cite{defEDS}.

An integral component of this EDS is the Intrusion Detection System (IDS). 
IDS analyzes network traffic to detect potential security breaches and produces logs corresponding to those activities \cite{Vigna1999}. An analyst can examine these logs and determine the next course of action. The analyst performs a variety of analytical and statistical operations to examine these logs. These may include charts and graphs. To enable these analyses, EDS provides the ability to compile and visualize these logs with the help of a Security Information and Event Management (SIEM) tool. This makes SIEM another very crucial component of the EDS.

IDS can be categorized into two primary types: signature-based and anomaly-based. Signature-based IDS relies on identifying known attack patterns and malicious signatures to generate alerts or take preventive actions \cite{Merve2021}. However, this approach often struggles to keep up with the rapid evolution of threats as it relies on predefined signatures.

Another approach employed by organizations to bolster their security is anomaly-based IDS. This approach analyzes network traffic to establish a baseline of normal behavior and then detects any deviations from the baseline as an indication of potential security risk \cite{Otoum2021}. Anomaly-based IDS offers the capability to detect previously unknown threats. Nevertheless, the anomaly-based detection approach is highly susceptible to generating many false positives and can lead to computational burden \cite{GARCIATEODORO200918}.

Due to the aforementioned limitations of these approaches, a hybrid-based IDS approach has gained prominence. This approach combines the strengths of both signature-based and anomaly-based IDS techniques \cite{Depren2005}. The combination of both these approaches provides a more comprehensive and robust defensive framework. Multiple research studies \cite{Depren2005, Yan2009, Abduvaliyev2010} have suggested that a hybrid-based IDS can achieve high detection rates while keeping false positives at a low level. Therefore, our research focuses on implementing a hybrid IDS framework within the EDS.

\section{Related Work}
In recent years, several studies have explored the integration of IDS and SIEM. Negoita and Carabas \cite{RW2} focused on enhancing security by integrating IDS with Machine Learning (ML) techniques using Elasticsearch. They used Snort \cite{snort} as IDS and leveraged Elasticsearch's built-in machine-learning framework for attack detection \cite{elk_ml}. Their study highlighted limitations in Elasticsearch's built-in ML jobs, such as manual configuration and difficulty detecting sophisticated attacks.

Priambodo et al. \cite{RW3} introduced an integration approach to enhance work-from-home network security. This approach combines Wireguard \cite{wireguard}, Suricata \cite{suricata} (an open-source IDS/IPS), and ELK (Elasticsearch \cite{elk}, Logstash \cite{logstash}, and Kibana \cite{kibana}). To evaluate the system's effectiveness, they generated port scanning attacks using Nmap \cite{nmap}. The detection of port scans and exploits was achieved through Suricata. The Kibana tool in the ELK server provided data log visualization for security hardening.

Esseghira et al. \cite{RW4} introduced the Aker security platform, which integrated IDS and SIEM functionalities and focused on analyzing encrypted network traffic. They used Suricata and Zeek \cite{zeek} as IDS, while Elasticsearch served as the SIEM system. They studied the growing prevalence of encrypted traffic and employed a decision-tree-based approach within Aker's threat investigation module. They assessed the effectiveness of Aker using User Acceptance Tests (UAT).

Muhammad et al. \cite{RW5} proposed using the ELK stack as an SIEM, Zeek as an IDS, and Slips \cite{slips} as a machine-learning analysis tool to build an integrated system. Their approach involved utilizing Slips for machine learning analysis of Zeek logs and forwarding the generated alerts to the ELK stack. They conducted simulations of DoS attacks to evaluate system performance by focusing on resource consumption metrics, such as CPU and RAM usage.

The existing literature primarily focuses on integrating open-source IDS tools with SIEM. Studies such as \cite{RW4} and \cite{RW5} utilize Zeek as the IDS, while others like \cite{RW2}, \cite{RW3}, and \cite{RW4} employ Suricata or Snort. Notably, some prior studies, including \cite{RW2} and \cite{RW5}, have applied methods or tools for anomaly detection. However, no prior study has proposed the integration of Suricata and Zeek as signature-based IDS, Slips as an anomaly-based IDS, and Elasticsearch as the SIEM platform within a single system. This research gap presents an opportunity to investigate the effectiveness of such a hybrid EDS.
\section{Methodology}
\begin{figure}[htbp]
\centering
\includegraphics[width=\linewidth]{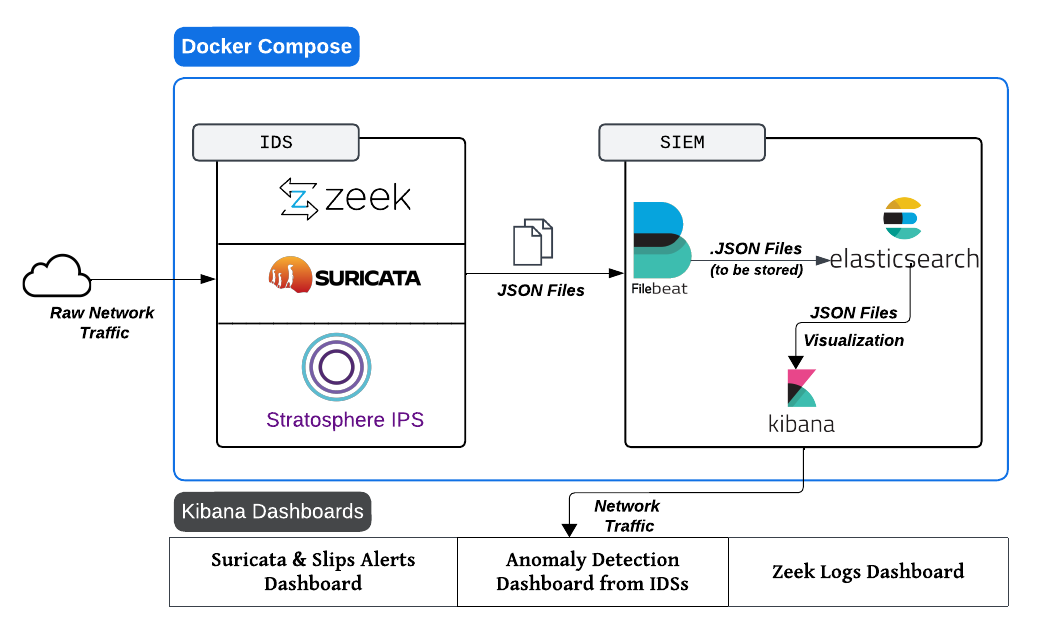}
\caption{Proposed EDS Architecture}
\label{fig:x EDS_Arch}
\end{figure}

The proposed EDS architecture, as illustrated in Figure \ref{fig:x EDS_Arch}, 
leverages three open-source IDSs: Zeek, Suricata, and Slips, and the SIEM solution, Elasticsearch. Zeek performs packet analysis, Suricata generates alert log files based on signature-based detection, and Slips generates alert log files based on anomaly-based detection.

Initially, we capture network traffic and send it to all three IDSs for analysis. Zeek, Suricata, and Slips examine the incoming traffic for malicious activities. We use Docker Compose \cite{docker} to deploy these IDS tools. With the help of Docker, we can easily deploy multiple containers of EDS within the network. 
Docker Compose further enhances the system's flexibility by enabling environment variables to dynamically pass crucial information like network interface, log paths, and Elasticsearch credentials across various IDS's YAML (or config) files during runtime. 

\begin{figure}[htbp]
\centering
\includegraphics[width=0.36\textwidth]{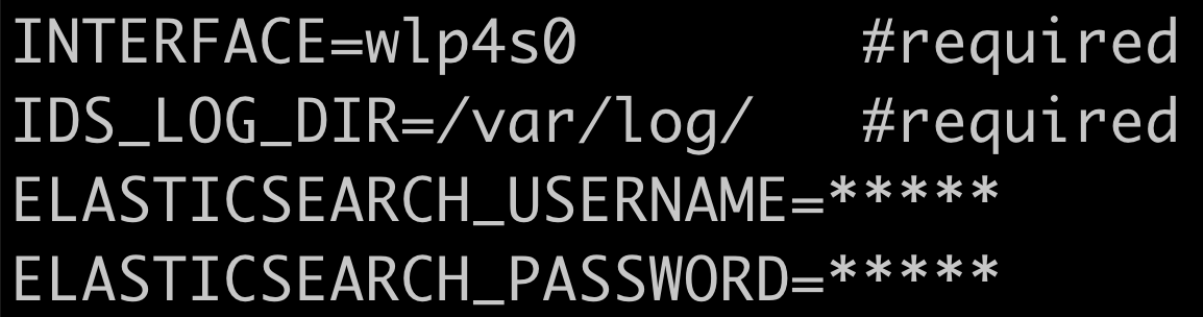}
\caption{EDS Configuration in the Docker .env File}
\label{fig:x edsEnv}
\end{figure}

Figure \ref{fig:x edsEnv} illustrates an example of environment variables
used in a Docker Compose file for the EDS. These variables
include:
\begin{itemize}
\item \verb|INTERFACE|: Specifies the host's network interface.
\item \verb|IDS_LOG_DIS|: Specifies the directory for log files.
\item \verb|ELASTICSEARCH_USERNAME_PASSWORD|: Specifies the Elasticsearch credentials.
\end{itemize}
Next, we send IDS alert logs to the SIEM solution, Elasticsearch. We use a lightweight data shipping utility called Filebeat \cite{filebeat} to perform this operation. It monitors the location of Zeek, Suricata, and Slips logs and sends them to Elasticsearch in JSON format. Upon receiving the logs, Elasticsearch indexes and stores them in its database, facilitating easy search and analysis \cite{elk}. Finally, we use Kibana to visualize the stored data. Kibana, a browser-based user interface, allows network administrators to filter, search, and display information in various formats \cite{kibana}.

\subsection{Intrusion Detection Systems (IDSs)}
The EDS incorporates three IDSs, namely Zeek, Suricata, and Slips, each offering unique features and capabilities.

\subsubsection{Zeek}
It is an intrusion detection tool that extracts files and metadata from network traffic, examines them, and consolidates them into its own alert mechanism \cite{zeek}. This alert mechanism contains customized logs created by Zeek highlighting insecure practices found in the network. \textit{conn.log}, produced by Zeek, plays a crucial role in monitoring network activity as it contains a list of insecure connections formed by the network.
 
\subsubsection{Suricata}
It provides signature-based threat detection methods. Suricata has an event information mechanism. All the events happening in the network are stored in the \textit{eve.json} file and forwarded to Elasticsearch\cite{suricataDocumentation}.
 
\subsubsection{Slips}
It uses ML-based anomaly detection to identify unknown attacks. Slips\cite{slips} analyzes the real-time network traffic and PCAP files and generates alerts in the form of \textit{alerts.json}, which are then sent to Elasticsearch for a thorough examination.

\setlength{\columnsep}{0.2 in}
\subsection{SIEM Solution}
Elasticsearch, an open-source search and analytics engine, is used in the EDS to store IDS log files \cite{elk}. Kibana is employed to visualize and interact with the data stored in Elasticsearch. To enhance data analysis in Kibana, we have designed customized dashboards focused on the Suricata and Slips alerts, an anomaly detection dashboard for all IDSs, and a dashboard for Zeek logs.

\section{Analysis and Result}
To evaluate the EDS, we have developed a bash script that enables a user to conduct cyber attacks on the network in a simulated environment.

\subsection{Attack Simulation Using Bash Script}
The primary objective of this script is to evaluate the EDS's effectiveness in attack identification. The script is available in the \textit{GitHub repository} \cite{github}. The script initiates by prompting the user to choose from a selection of attack tools: a) Nmap, b) Nikto \cite{nikto}, c) Ping \cite{ping}, d) Hping \cite{hping}, and e) SQLMap \cite{sqlmap}.

While these attacks are executed, the IDSs continuously run in the background to detect intrusive behavior. The IDS logs generated during these scenarios are sent to Elasticsearch for analysis. To filter malicious logs on Kibana, Kibana Query Language (KQL) \cite{kibana} has been used for customized visualization.

\subsection{Detection of Attacks Utilizing Elasticsearch}
This section will explore how the bash script works for each attack. We will also discuss how Elasticsearch and Kibana can be customized for each attack scenario.

\subsubsection{Port Scanning}
We utilized the following tools to scan the network. Nmap: \verb|nmap -sS <ip> -p 1-1000|, Ping: \verb|ping -c 10 <ip>|, and Nikto: \verb|nikto -h <ip>|.

To visualize these attacks on Kibana, we used the following KQL query: 
\begin{verbatim}
not (network.direction: "outbound") 
and ((not (network.transport: "icmp") and 
not(zeek.connection.history:/Sh*|F*|D*/))
or (network.transport: "icmp" 
and zeek.connection.icmp.type: "8"))
\end{verbatim}

This KQL query identifies port scanning attempts by Nmap and Nikto by filtering network traffic. First, it excludes outbound traffic and then applies two alternative conditions. The first condition is that the query filters network traffic that does not use the ICMP transport protocol and does not match the specified patterns in the Zeek connection history. This can help identify TCP/UDP port scanning attempts. The second condition filters out traffic using the ICMP transport protocol and containing an ICMP type of "8" (echo request). This can help identify ICMP-based port scan attempts (ping scan).

\begin{figure}[htbp]
\centering
\includegraphics[width=0.47\textwidth]{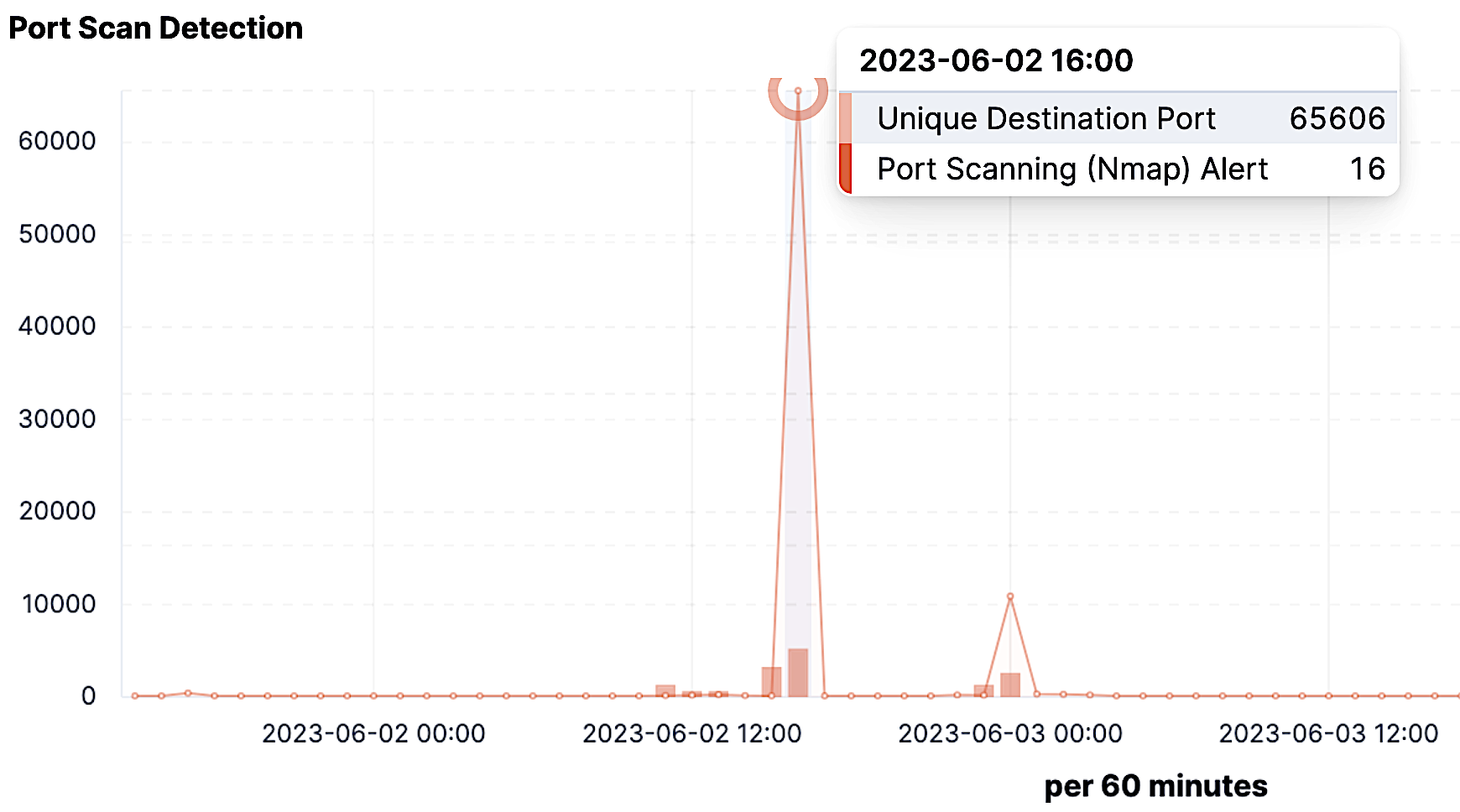}
\caption{Port Scanning Attack Detection Using KQL}
\label{fig:x NmapAndItsPing}
\end{figure}

In Figure \ref{fig:x NmapAndItsPing}, the graph illustrates port scanning attempts, represented by the shaded light red region beneath the straight line graph. These attempts were predominantly caused by running Nmap on the network. The red bars in the graph indicate the number of port-scanning attack alerts generated by Suricata and Slips. These alerts are filtered using KQL to identify alerts related to attempted information leaks from Suricata and reconnaissance scanning from Slips.

\subsubsection{Denial-of-Service (DoS)} To simulate DoS attacks, we utilized the Hping tool with the command: \verb|hping3 -c 100 -p 21 -w 64 -d 120 --flood| \verb|--rand-source <ip>|. This command floods the target with a high volume of packets. Attacks were launched on ports 21 and port 80. 

The KQL query used for detecting DoS attacks is:
\begin{verbatim}
 not (network.direction: "outbound") 
\end{verbatim}
This query excludes outbound traffic, focusing on inbound or internal traffic to narrow down the visualization in the graph and detect potential DoS attacks.

\begin{figure}[htbp]
\centering
\includegraphics[width=0.47\textwidth]{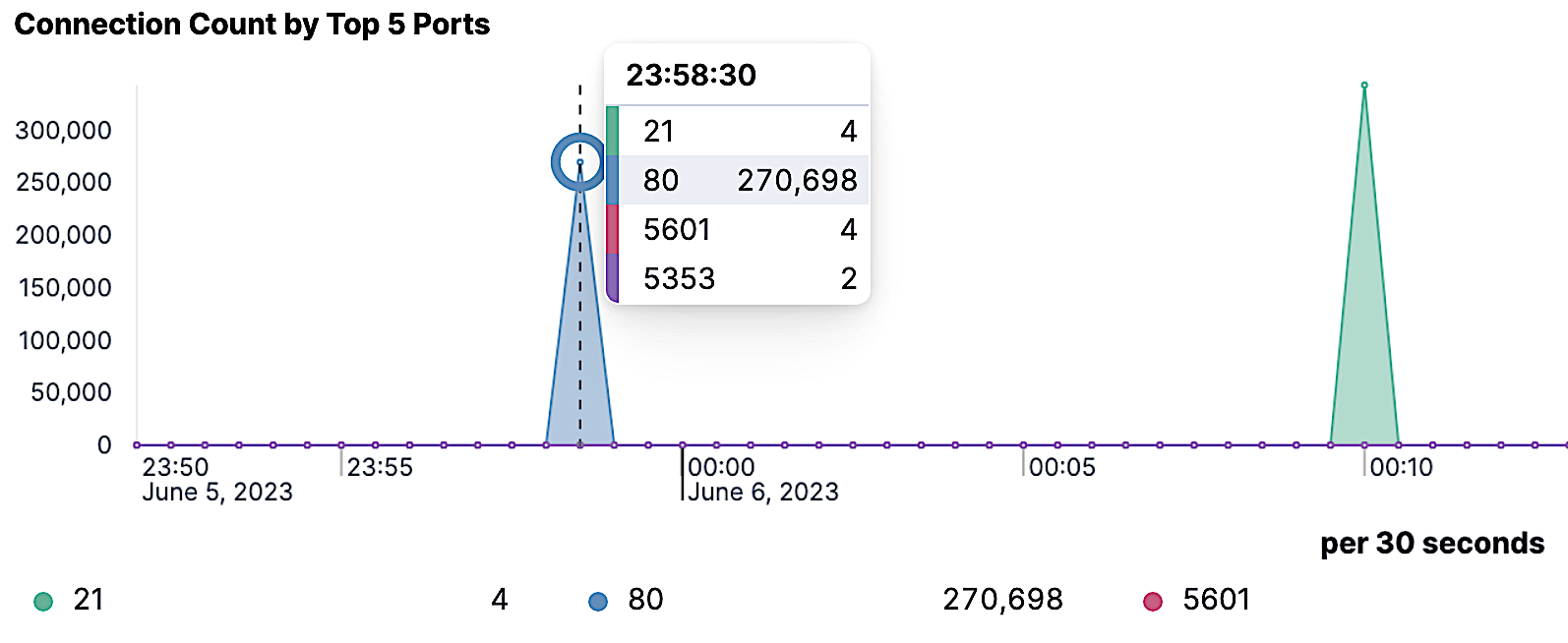}
\includegraphics[width=0.47\textwidth]{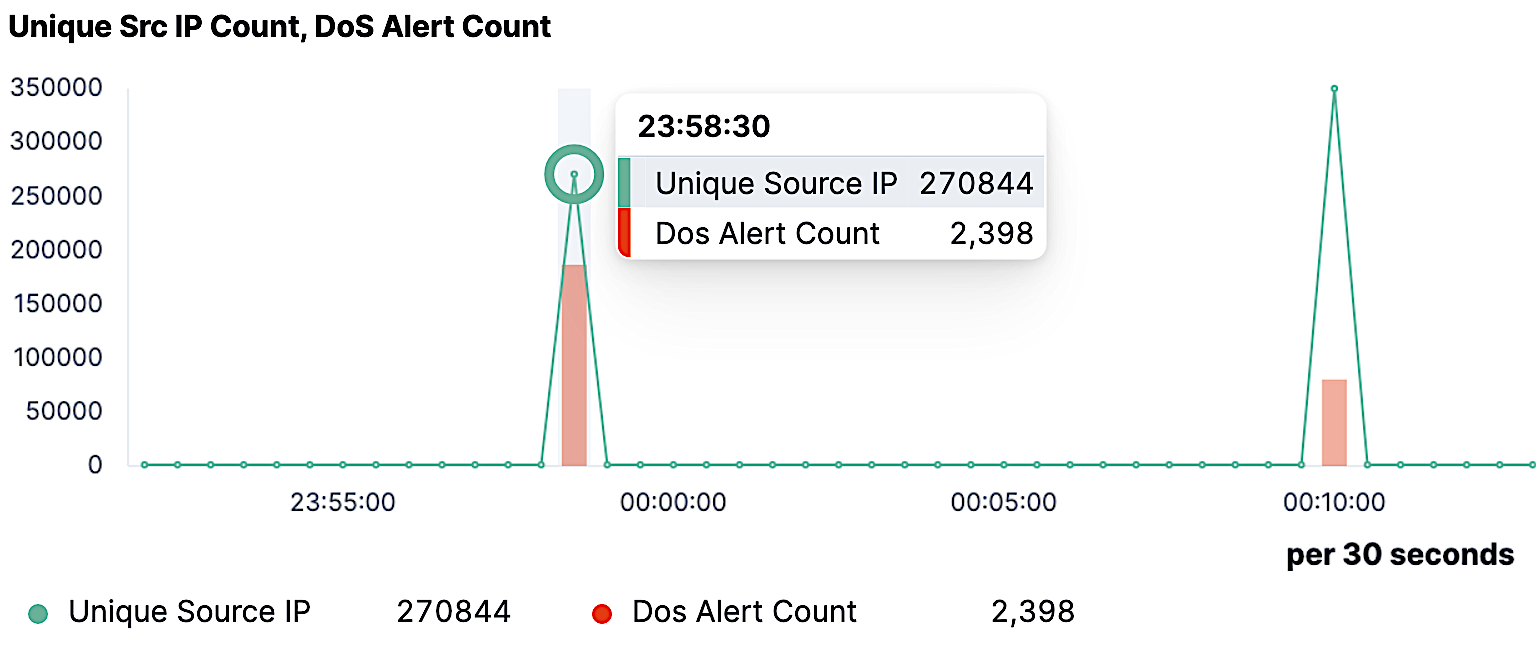}
\caption{DoS Attack Detection Using KQL}
\label{fig:x hping3}
\end{figure}

In Figure \ref{fig:x hping3}, the graph on top illustrates the targeted port numbers for the DoS attacks. The shaded green region represents the DoS attack on port 21, while the blue shade represents the same attack on port 80. In the graph on the bottom, the red bars depict the total number of alerts generated by Suricata and Slips.

During the DoS attack simulations, 270,717 packets were transmitted through port 80, resulting in 2,398 alerts generated by Suricata and Slips. For port 21, 214,224 packets were transmitted, and both Suricata and Slips generated a total of 670 alerts. These statistics provide valuable insights into the volume of network traffic observed during a DoS attack and demonstrate the EDS's capability to detect and visualize these alerts.

\subsubsection{Privilege Escalation}
To evaluate the EDS's capability in detecting privilege escalation attacks, we utilized the SQLMap tool, commonly utilized for SQL injection (SQLi) attacks. To visualize the attack, we implemented the following KQL query:

\begin{verbatim}
 user_agent.original: sqlmap* 
\end{verbatim}

It filters network logs to specifically search for user agent strings containing the term "sqlmap." This query allows us to capture and analyze the network traffic associated with sqlmap requests.

\begin{figure}[htbp]
\centering
\includegraphics[width=0.47\textwidth]{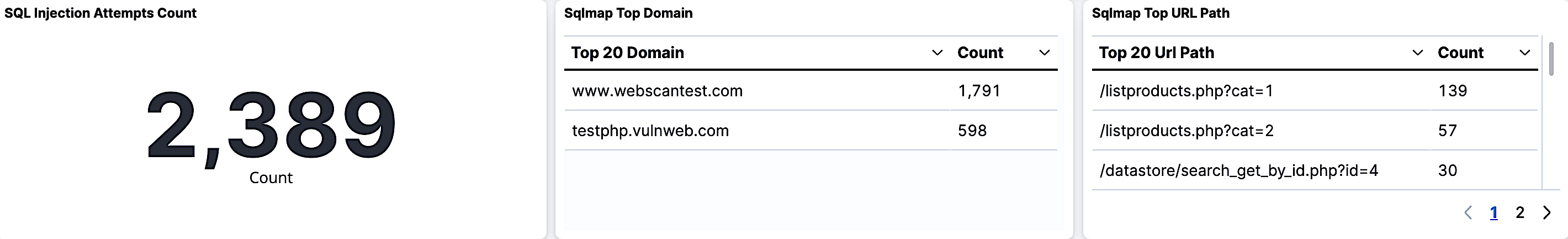}
\includegraphics[width=0.47\textwidth]{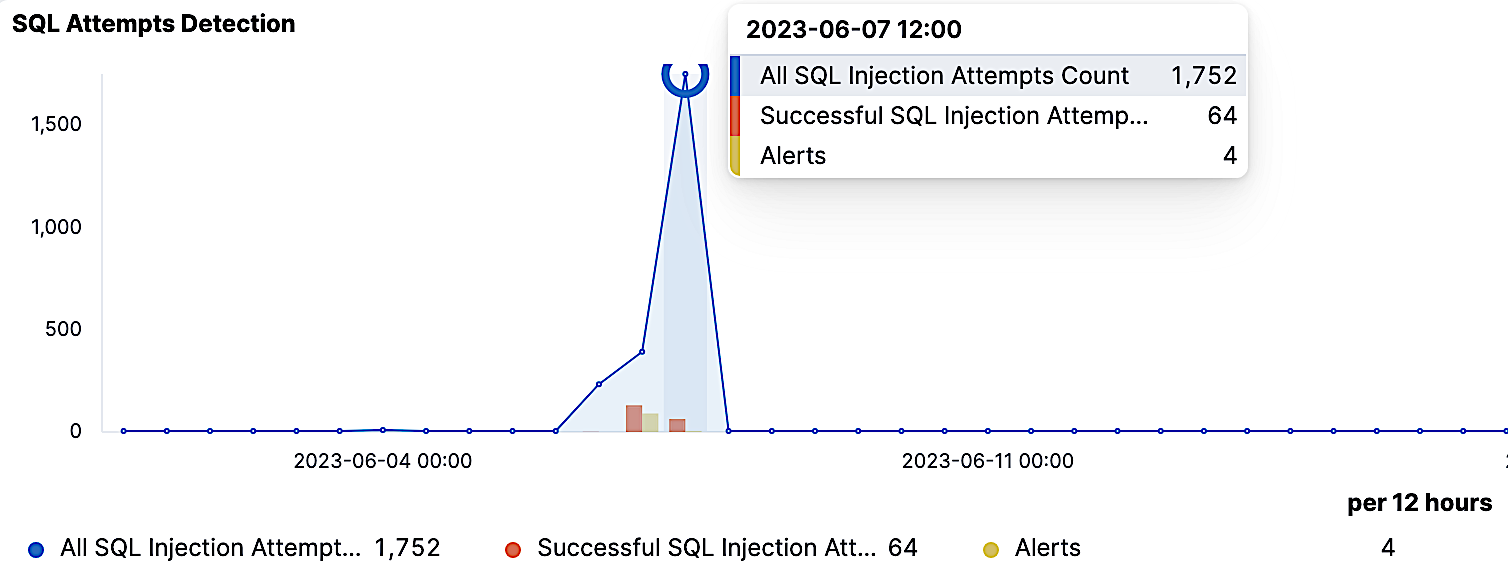}
\caption{SQLi Attack Detection Using KQL}
\label{fig:x mysqlDetection}
\end{figure}

Figure \ref{fig:x mysqlDetection} shows the domain name and complete URL of the attack, providing valuable insights into the attacker's methodology. Furthermore, the graph indicates the EDS's ability to detect potential SQLi attacks. The shaded light blue region beneath the straight line graph represents the total number of SQLi attacks, the red bar represents the number of successful SQLi attacks, and the yellow bar represents the number of alerts generated by Suricata and Zeek.

\section{Conclusion}
EDS can enhance network security by integrating IDSs with SIEM to provide a comprehensive defense solution. The study evaluates the efficacy of integrating signature-based and anomaly-based IDSs. Additionally, the leverage of SIEM enhances the EDS with user-friendly interfaces, facilitating efficient threat detection. The research evaluates the EDS's robustness in detecting various threats like port scanning, DoS, and privilege escalation. The outcomes of this research hold implications for the field of network security and contribute to strengthening cyber defense strategies.

\bibliographystyle{IEEEtran}
\bibliography{main}

\end{document}